\documentclass[useAMS,usenatbib,onecolumn,usegraphicx]{mn2e}
\usepackage{color}
\usepackage{amssymb}
\newcommand{\beq}{\begin{equation}}
\newcommand{\eeq}{\end{equation}}
\newcommand{\beqar}{\begin{eqnarray}}
\newcommand{\eeqar}{\end{eqnarray}}
\newcommand{\beqars}{\begin{eqnarray*}}
\newcommand{\eeqars}{\end{eqnarray*}}
\newcommand{\bc}{\begin{center}}
\newcommand{\ec}{\end{center}}
\newcommand{\ben}{\begin{enumerate}}
\newcommand{\een}{\end{enumerate}}
\newcommand{\bit}{\begin{itemize}}
\newcommand{\eit}{\end{itemize}}
\def \non{\nonumber}

\def \veps{\varepsilon}
\renewcommand{\(}{\left(}
\renewcommand{\)}{\right)}

\begin{document}

\title[Crustal Oscillations of Slowly Rotating Relativistic Stars]
      {Crustal Oscillations of Slowly Rotating Relativistic Stars}
\author[M.~Vavoulidis, K.~D.~Kokkotas, A.~Stavridis]
{M.~Vavoulidis$^1$ \thanks{E-mail: miltos@astro.auth.gr}
, K.~D.~Kokkotas$^{1,2}$ and A.~Stavridis$^1$ \\
$^1$ Department of Physics, Aristotle University of Thessaloniki, 54124, Thessaloniki, Greece \\
$^2$ Theoretical Astrophysics, Eberhard-Karls University of T\"ubingen, 72076, T\"ubingen, Germany}


\maketitle


\begin{abstract}
We study low-amplitude crustal oscillations of slowly rotating relativistic stars
consisting of a central fluid core and an outer thin solid crust.
We estimate the effect of rotation
on the torsional toroidal modes
and on the interfacial and shear spheroidal modes.
The results compared against the Newtonian ones
for  wide range of neutron star models and equations of state.
\end{abstract}


\section{Introduction}

Crustal oscillations of nonrotating nonmagnetic neutron stars
have been studied in Newtonian theory
\citep{HC1980,McDermottEtAl1985,MVHH1988,StrohmayerEtAl1991,BastrukovEtAl2007}
as well as in General Relativity
\citep{ST1983,Finn1990,Leins1994,YL2002,SA2007}.
These studies have been gradually extended in order to include
the effects of rotation \citep{Strohmayer1991,LS1996,Lee2007a,VSKB2007}
or strong magnetic fields
\citep{CarrollEtAl1986,Duncan1998,MPS2001,Piro2005,Lee2007a,SKS2007a,Lee2007b,SCK2007}.
Most of them have been focused on torsional toroidal modes of oscillation (designated as $_\ell t_n$)
and only a few studies have dealt with
interfacial ($_\ell i$) and shear ($_\ell s_n$) spheroidal modes. There are also some recent studies of the perturbations of purely elastic stars in the general relativistic framework \citep{Karlovini2004,Karlovini2007}.

After the discovery
of high-frequency quasi-periodic oscillations (QPOs)
in the tails of giant flares from soft gamma-ray repeaters (SGRs)
\citep{BaratEtAl1983,IsraelEtAl2005,SW2005,WS2006,SW2006},
special attention has been drawn to
torsional toroidal oscillations of neutron star crusts.
Most of the QPOs have been observed at frequencies between 18 and 155 Hz,
although there have been a few at  higher frequencies,
for example at 625, 1840 and possibly at 718 Hz for SGR 1806-20
\citep{WS2007}.
Fundamental torsional modes could account for many of the low frequencies \citep{Duncan1998}
while radial overtones could account for some of the higher ones \citep{Piro2005}.
However, the torsional-mode interpretation had two drawbacks:
it could not explain the very low observed frequency at 18 Hz and
it could not explain observed pairs of frequencies
as those at 26 and 30 Hz or those at 625 and 718 Hz.

It has been soon realized
that the coupling between the crust and the magnetic field
would play a key-role in our attempt to explain the observed QPOs.
In fact, this coupling would favour the existence of global magnetoelastic modes of oscillation
rather than pure elastic modes confined in the crust \citep{Levin2006}.
Moreover, it has been argued
that these modes should decay on a short timescale
because of the presence of a magnetohydrodynamical continuum in the core
and only specific QPOs could be long-lived
\citep{GSA2006,SKSV2006,Lee2007a,Levin2007,SKS2007b}.

After a catastrophic reconfiguration of the stellar magnetic field,
axial-type torsional and Alfv\'en oscillations should be the most easily excited
as polar-type oscillations would have to overcome strong restoring forces.
However, the presence of a magnetic field would inevitably couple these oscillations
with polar-type ones characterized by interfacial (i-), shear (s-),
pressure-restored (f- and p-), gravity (g-) and polar-type magnetoelastic modes;
stellar rotation should do more or less the same.
These polar-type modes would involve density variations
and could be relevant for gravitational-wave emission \citep{LIGO2007}.
The possible association of polar-type modes with the observed QPOs
has been hinted by \citet{Piro2005,WS2006}
but further investigation is definitely needed.

In Section \ref{Formulation} we derive the equations
that describe linear spheroidal and toroidal oscillations of slowly rotating relativistic stars
possessing a crust.
In most of them,
a parameter $\veps$ indicates the presence of a relativistic term.
Therefore $\veps \rightarrow 0$ becomes our standard tool to recover the already known Newtonian
equations for these oscillations \citep{Strohmayer1991,LS1996}.
Furthermore, the nonrotating parts of these relativistic equations
are easily comparable with the equations given in \cite{YL2002}.
In Section \ref{Formulation} we derive the perturbation equations for the solid-crust region (\ref{SolidCrust}),
we review those for the fluid-core region (\ref{FluidCore})
and we present the necessary physical boundary and jump conditions (\ref{BNConditions}).
In Section \ref{Results} we present our numerical results for a set of neutron star models
with different equations of state (EoS) and different bulk properties (\ref{NSmodels}),
first revisiting torsional toroidal modes (\ref{Torsional_modes})
and then focusing on interfacial and shear spheroidal modes (\ref{Interfacial_shear_modes}).
In Section \ref{Discussion} we summarize and discuss our results.

\section{Formulation} \label{Formulation}

We consider a slowly rotating relativistic, strain free, star described by the metric:
\beq
ds^2 = - e^{2\nu} dt^2 + e^{2\lambda} dr^2 + r^2 d\theta^2 + r^2 \sin^2\theta d\phi^2 - 2 \omega r^2 \sin^2\theta dt d\phi ,
\eeq
where $\nu$, $\lambda$ and $\omega$ are functions of the radial coordinate $r$.
These functions are solutions of the Tolman-Oppenheimer-Volkoff (TOV) equations:
\beqar
e^{2\lambda} &=& \left(1 - \veps \frac{2M\left(r\right)}{r} \right)^{-1} ,
\; \; \; \frac{dM\left(r\right)}{dr} = 4 \pi r^2 \rho , \label{TOV1} \\
\frac{d\nu}{dr} &=& e^{2\lambda} \left(4 \pi r \veps p + \frac{M\left(r\right)}{r^2} \right) ,
\; \; \; \frac{dp}{dr} = - \left(\rho+\veps p\right) \frac{d\nu}{dr} , \\
\frac{d\lambda}{dr} &=& \veps e^{2\lambda} \left(4 \pi r \rho - \frac{M\left(r\right)}{r^2} \right) , \label{TOV3}
\eeqar
and of one more equation
describing the dragging of the inertial frames of reference \citep{Hartle1967}:
\beq
\frac{d^2\varpi}{dr^2} - \left(\frac{d\nu}{dr} + \frac{d\lambda}{dr} - \frac{4}{r} \right) \frac{d\varpi}{dr}
- 16 \pi e^{2\lambda} \left(\rho+p\right) \varpi = 0 ,
\eeq
where $\varpi := \Omega - \omega$,
$\rho$ and $p$ are the energy density and the pressure, respectively,
$M\left(r\right)$ is the mass inside radius $r$
and $\Omega$ is the stellar rotational frequency.
In our general-relativistic approach,
the parameter $\veps$ equals to 1.
Its presence aims to point out
the Newtonian limit where $\veps \rightarrow 0$.
Then, according to equations (\ref{TOV1})-(\ref{TOV3}),
$\lambda \rightarrow 0$,
$d\nu/dr \rightarrow M\left(r\right)/r^2$,
$dp/dr \rightarrow - \rho d\nu/dr$
and
$d\lambda/dr \rightarrow 0$.
Furthermore, in the Newtonian limit,
$\omega \rightarrow 0$ or, equivalently, $\varpi \rightarrow \Omega$ and $d\varpi/dr \rightarrow 0$.

The pulsation equations
come from the linear perturbation of the energy-momentum conservation law
$\delta \left(\nabla^\beta T_{\alpha \beta} \right) = 0$
where:
\beq
T_{\alpha \beta} = \left(\rho+p\right) u_\alpha u_\beta + p g_{\alpha \beta} - 2 \mu S_{\alpha \beta} . \label{Tab}
\eeq
The components of the perturbed four-velocity are given by the relation $\delta u^\alpha = {\cal L}_u \xi^\alpha$
where $\xi^\alpha$ is the displacement vector
while the shear tensor $S_{\alpha \beta}$ is given by the relation $\sigma_{\alpha \beta} = {\cal L}_u S_{\alpha \beta}$
where $\sigma_{\alpha \beta}$ is the rate of shear tensor. Namely $\sigma_{\alpha \beta}$ is the Lie derivative
of the shear tensor along the world lines
\citep{CQ1972}
and is calculated by the equation:
\begin{equation}
\sigma_{\alpha \beta} = {1 \over 2} \(
P^\gamma_\beta \nabla_\gamma u_\alpha +
P^\gamma_\alpha \nabla_\gamma u_\beta \)
- { 1 \over 3} P_{\alpha \beta} \nabla_\gamma u^\gamma ,
\end{equation}
where $P_{\alpha \beta}$ is the projection tensor:
\begin{equation}
P_{\alpha \beta} = g_{\alpha \beta} + u_{\alpha} u_{\beta} . \label{Pab}
\end{equation}
It is implied that we work in the Cowling approximation
as we neglect the perturbed Einstein equations $ \delta G_{\alpha
  \beta}=8\pi  \delta T_{\alpha \beta}$
and, additionally, set all metric perturbations equal to zero
in our equations. The Cowling approximation is typically very good for toroidal type of oscillations i.e. $t$-modes, $r$-modes etc, while typically for polar type of perturbations the error can be of the order of 10-20\% especially for the fundamental pressure mode the $f$-mode. For, the other type of modes, like $g$-modes, higher $p$-modes and $s$-modes one does not expect deviations larger than 3-5\%. 
  
We choose to work in a corotating reference frame
where $\delta u^\alpha = e^{-\nu}\partial \xi^\alpha/\partial t$,
$\sigma_{\alpha \beta} = e^{-\nu} \partial S_{\alpha \beta}/\partial t$.
Using equations (\ref{Tab})-(\ref{Pab}),
the energy-momentum conservation law yields
three second-order partial differential equations
for the three components of the displacement vector
$\xi^r=\xi^r\left(t,r,\theta,\phi\right)$,
$\xi^\theta=\xi^\theta\left(t,r,\theta,\phi\right)$ and
$\xi^\phi=\xi^\phi\left(t,r,\theta,\phi\right)$.
These equations are the relativistic version
of the Newtonian equations (35)-(39) of \cite{Strohmayer1991}
and their explicit form is given in the Appendix \ref{StrohmayerGR}.

When we focus on spheroidal modes, we can write for the displacement vector:
\beq
\xi^i
= \left[rS, H \frac{\partial}{\partial \theta}, H \frac{1}{\sin^2\theta} \frac{\partial}{\partial \phi}\right] Y_{\ell m} e^{i\sigma t} ,
\label{xi_spheroidal}
\eeq
while when we focus on toroidal modes, we can write:
\beq
\xi^i = \left[0, T \frac{1}{\sin\theta} \frac{\partial}{\partial \phi},
- T \frac{1}{\sin\theta} \frac{\partial}{\partial \theta} \right] Y_{\ell m} e^{i\sigma t} .
\eeq
Working in the slow-rotation approximation
\citep{Kojima1992,SK2005},
we eventually obtain
three second-order ordinary differential equations for the functions $S=S\left(r\right), H=H\left(r\right)$ and $T=T\left(r\right)$.
The equations for $S$ and $H$ are quite lengthy and we do not show them here. However, we mention
that they are the relativistic version of the equations used in \cite{LS1996}. On the other hand,
the equation for the toroidal radial function $T$, being shorter, reads \citep{VSKB2007}:
\beq
- \sigma^2 T = v_s^2 e^{2\veps\left(\nu-\lambda\right)}
\left[\frac{d^2 T}{dr^2}
+\left(\frac{4}{r} + \veps \frac{d\nu}{dr} - \veps \frac{d\lambda}{dr} + \frac{1}{\mu} \frac{d\mu}{dr} \right) \frac{dT}{dr}
- e^{2\veps\lambda} \frac{\Lambda-2}{r^2} T \right]
- 2 m \sigma \varpi \left[\frac{1}{\Lambda}+\veps v_s^2\left(1-\frac{2}{\Lambda}\right)\right] T ,
\label{T}
\eeq
where $v_s^2 := \mu/\left(\rho+\veps p\right)$ is the speed of shear waves and $\Lambda := \ell \left(\ell+1\right)$.

We then expand the frequency and the displacement functions in power series:
\beq
\sigma = \sum_{j=0}^{\infty} \eta^j \sigma_j = \sigma_0 + \eta \sigma_1 + {\cal O} \left(\eta^2\right), \; \; \;
\xi^i = \sum_{j=0}^{\infty} \eta^j \xi^{i,j} = \xi^{i,0} + \eta \xi^{i,1} + {\cal O} \left(\eta^2\right) ,
\eeq
where $\eta$ is an auxiliary expansion parameter
and we insert these expansions into the main perturbation equations, e.g. equation (\ref{T}) above.
In this way we split each equation in two,
one zeroth-order in $\Omega$
which is taken collecting all $\eta^0$ terms
and one first-order in $\Omega$
which is taken collecting all $\eta^1$ terms.
For example, equation (\ref{T}), to zeroth order in rotation, gives the equation:
\beq
- \sigma_0^2 T^0 = v_s^2 e^{2\veps\left(\nu-\lambda\right)}
\left[\frac{d^2 T^0}{dr^2}
+\left(\frac{4}{r} + \veps \frac{d\nu}{dr} - \veps \frac{d\lambda}{dr} + \frac{1}{\mu} \frac{d\mu}{dr} \right)
\frac{dT^0}{dr}
- e^{2\veps\lambda} \frac{\Lambda-2}{r^2} T^0 \right] , \label{Subequation1}
\eeq
which, when supplied with the appropriate boundary conditions,
determines the zeroth-order eigenfrequency, $\sigma_0$, and the zeroth-order eigenfunction, $T^0$.
Then, to first order in rotation, equation (\ref{T}) gives another equation:
\beqar
- \sigma_0^2 T^1 - 2 \sigma_0 \sigma_1 T^0 &=& v_s^2 e^{2\veps\left(\nu-\lambda\right)}
\left[\frac{d^2 T^1}{dr^2}
+\left(\frac{4}{r} + \veps \frac{d\nu}{dr} - \veps \frac{d\lambda}{dr} + \frac{1}{\mu} \frac{d\mu}{dr} \right)
\frac{dT^1}{dr}
- e^{2\veps\lambda} \frac{\Lambda-2}{r^2} T^1 \right]
\non
\\
&-& 2 m \sigma_0 \varpi \left[\frac{1}{\Lambda}+\veps v_s^2\left(1-\frac{2}{\Lambda}\right)\right] T^0 , \label{Subequation2}
\eeqar
which determines the first-order rotational corrections of the eigenfrequency, $\sigma_1$,
and of the eigenfunction, $T^1$,
using the known zeroth-order quantities $\sigma_0$ and $T^0$,
already calculated by equation (\ref{Subequation1}).

\subsection{Solid crust} \label{SolidCrust}

To describe the oscillations in the solid-crust region, we use the following functions:
\beqar
z_1^j &=& S^j , \\
z_2^j &=& 2 \alpha_1 e^{-\veps\lambda} \frac{d}{dr} \left(r e^{\veps\lambda} S^j\right)
+\left(\Gamma-\frac{2}{3}\alpha_1\right)
\left\{\frac{e^{-\veps\lambda}}{r^2} \frac{d}{dr} \left(r^3 e^{\veps\lambda} S^j\right)
- \ell\left(\ell+1\right) H^j \right\} , \\
z_3^j &=& H^j , \\
z_4^j &=& \alpha_1 \left(e^{-2\veps\lambda} r \frac{dH^j}{dr} + S^j \right) , \\
z_5^j &=& T^j , \\
z_6^j &=& \alpha_1 e^{-2\veps\lambda} r \frac{dT^j}{dr} .
\eeqar
This set of functions is the set (31)-(36) of \cite{YL2002}
and it reduces to the set of functions (53),(16),(15) of \cite{LS1996} in the Newtonian limit, $\veps\rightarrow 0$.

Using these functions, we can recast our three main perturbation equations
into a system of six first-order ordinary differential equations
where four of them are describing the spheroidal perturbations
while the other two are describing the toroidal perturbations.
To zeroth order in $\Omega$ $\left(j=0\right)$, this system has the form:
\beqar
r \frac{dz_1^0}{dr} &=&
- \left(1+2\frac{\alpha_2}{\alpha_3}+\veps U_2\right) z_1^0
+ \frac{1}{\alpha_3} z_2^0
+ \frac{\alpha_2}{\alpha_3} \ell\left(\ell+1\right) z_3^0 , \label{rdz10dr} \\
r \frac{dz_2^0}{dr} &=&
\left\{\left(-3-\veps U_2+U_1-e^{2\veps\lambda}c_1\bar{\sigma}_0^2\right) V_1
+ 4\frac{\alpha_1}{\alpha_3}\left(3\alpha_2+2\alpha_1\right)
\right\} z_1^0
+ \left(V_2-4\frac{\alpha_1}{\alpha_3}\right) z_2^0
\non
\\
&+& \left\{V_1 - 2 \alpha_1 \left(1+2\frac{\alpha_2}{\alpha_3}\right)
\right\} \ell \left(\ell+1\right) z_3^0
+ e^{2\veps\lambda} \ell \left(\ell+1\right) z_4^0 , \\
r \frac{dz_3^0}{dr} &=&
-e^{2\veps\lambda} z_1^0 + \frac{e^{2\veps\lambda}}{\alpha_1} z_4^0 , \\
r \frac{dz_4^0}{dr} &=&
- \left(-V_1+6\Gamma\frac{\alpha_1}{\alpha_3} \right) z_1^0
- \frac{\alpha_2}{\alpha_3} z_2^0
- \left\{c_1 \bar{\sigma}_0^2 V_1 + 2\alpha_1 -2 \frac{\alpha_1}{\alpha_3}
\left(\alpha_2+\alpha_3\right)\ell\left(\ell+1\right) \right\} z_3^0
- \left(3+\veps U_2-V_2\right) z_4^0 , \\
r \frac{dz_5^0}{dr} &=&
\frac{e^{2\veps\lambda}}{\alpha_1} z_6^0 , \\
r \frac{dz_6^0}{dr} &=&
-\left\{c_1 \bar{\sigma}_0^2 V_1 - \alpha_1 \left(\ell-1\right)\left(\ell+2 \right) \right\} z_5^0
- \left(3 + \veps U_2 - V_2 \right) z_6^0 . \label{rdz60dr}
\eeqar
This set of equations is the set (25)-(30) of \cite{YL2002}
and, in the Newtonian limit,
reduces to the sets of equations (54)-(57) and (68)-(69) of \cite{LS1996}.
The first four equations give the zeroth-order eigenfrequency $\sigma_0$ and eigenfunctions
$z_1^0, z_2^0, z_3^0, z_4^0$ for the interfacial and shear spheroidal oscillations
while the last two equations give $\sigma_0, z_5^0$ and $z_6^0$ for the torsional toroidal oscillations.

As in \cite{YL2002}, the functions that appear in equations (\ref{rdz10dr})-(\ref{rdz60dr}) are:
\beqar
\alpha_1 = \frac{\mu}{p}, \; \; \; \alpha_2 = \Gamma - \frac{2}{3} \alpha_1, \; \; \; \alpha_3 = \Gamma + \frac{4}{3} \alpha_1,
\\
V_1 = \frac{\rho+\veps p}{p} r \frac{d\nu}{dr}, \; \; \; V_2 = \frac{\rho}{p} r \frac{d\nu}{dr},
\label{Vcoeffs}
\\
U_1 = \left(\frac{d\nu}{dr}\right)^{-1} \frac{d}{dr} \left(r \frac{d\nu}{dr} \right), \; \; \; U_2 = \veps r \frac{d\lambda}{dr},
\\
c_1 = \frac{M}{R^3} r e^{-2\veps\nu} \left(\frac{d\nu}{dr}\right)^{-1} .
\label{Ccoeff}
\eeqar
In the Newtonian limit,
$\veps\rightarrow 0$ and
$d\nu/dr\rightarrow g$ where $g = M\left(r\right)/r^2$ is the gravitational acceleration
with $dp/dr \rightarrow -\rho g$.
Therefore, the functions (\ref{Vcoeffs})-(\ref{Ccoeff}) reduce to:
\beqar
V_1, V_2 \rightarrow \frac{\rho}{p} rg = - \frac{d\ln p}{d \ln r} = V,
\\
U_1 \rightarrow g^{-1} \frac{d}{dr} \left(rg\right) = \frac{r^2}{M\left(r\right)} \frac{d}{dr} \left(\frac{M\left(r\right)}{r}\right)
= \frac{d\ln M\left(r\right)}{d\ln r} - 1 = U - 1 ,
\\
c_1 \rightarrow \frac{M}{R^3} r g^{-1} = \left(\frac{r}{R}\right)^3 \frac{M}{M\left(r\right)} .
\eeqar
The functions $V,U$ and $c_1$ are well-known from the Newtonian theory,
see e.g. the set of relations (35) of \cite{LS1996} or the set of relations (15) of \cite{MVHH1988}.

To first order in $\Omega$ $\left(j=1\right)$, our system of equations has the form:
\beqar
r \frac{dz_1^1}{dr} &=&
- \left(1+2\frac{\alpha_2}{\alpha_3}+\veps U_2\right) z_1^1
+ \frac{1}{\alpha_3} z_2^1
+ \frac{\alpha_2}{\alpha_3} \ell\left(\ell+1\right) z_3^1 ,
\label{rdz11dr} \\
r \frac{dz_2^1}{dr} &=&
\left\{\left(-3-\veps U_2+U_1-e^{2\veps\lambda}c_1\bar{\sigma}_0^2\right) V_1
+ 4\frac{\alpha_1}{\alpha_3}\left(3\alpha_2+2\alpha_1\right)
\right\} z_1^1
+ \left(V_2-4\frac{\alpha_1}{\alpha_3}\right) z_2^1
\non
\\
&+& \left\{V_1 - 2 \alpha_1 \left(1+2\frac{\alpha_2}{\alpha_3}\right)
\right\} \ell \left(\ell+1\right) z_3^1
+ e^{2\veps\lambda} \ell \left(\ell+1\right) z_4^1
\non
\\
&+& \left\{- 2 e^{2\veps\lambda}c_1 \bar{\sigma}_0 \bar{\sigma}_1 V_1 + \veps {\cal A} \right\} z_1^0
+ \left\{  2 m c_1 \bar{\sigma}_0 \bar{\varpi} V_1 + \veps {\cal B} \right\} z_3^0
+ \veps {\cal C} z_4^0
, \\
r \frac{dz_3^1}{dr} &=&
-e^{2\veps\lambda} z_1^1 + \frac{e^{2\veps\lambda}}{\alpha_1} z_4^1 , \\
r \frac{dz_4^1}{dr} &=&
- \left(-V_1+6\Gamma\frac{\alpha_1}{\alpha_3} \right) z_1^1
- \frac{\alpha_2}{\alpha_3} z_2^1
- \left\{c_1 \bar{\sigma}_0^2 V_1 + 2\alpha_1 -2 \frac{\alpha_1}{\alpha_3}
\left(\alpha_2+\alpha_3\right)\ell\left(\ell+1\right) \right\} z_3^1
- \left(3+\veps U_2-V_2\right) z_4^1
\non
\\
&+& \left\{\frac{2 m c_1 \bar{\sigma}_0 \bar{\varpi} V_1}{\ell\left(\ell+1\right)} + \veps {\cal D} \right\} z_1^0
+ \veps {\cal E} z_2^0
+ \left\{- 2 c_1 \bar{\sigma}_0 \bar{\sigma}_1 V_1 + \frac{2 m c_1 \bar{\sigma}_0 \bar{\varpi} V_1}{\ell\left(\ell+1\right)}
+ \veps {\cal F} \right\} z_3^0 ,
\label{rdz41dr}
\\
r \frac{dz_5^1}{dr} &=&
\frac{e^{2\veps\lambda}}{\alpha_1} z_6^1 , \\
r \frac{dz_6^1}{dr} &=&
-\left\{c_1 \bar{\sigma}_0^2 V_1 - \alpha_1 \left(\ell-1\right)\left(\ell+2 \right) \right\} z_5^1
- \left(3 + \veps U_2 - V_2 \right) z_6^1
\non
\\
&+& \left\{2 c_1 \bar{\sigma}_0 V_1 \left[- \bar{\sigma}_1 + \frac{m \bar{\varpi}}{\ell\left(\ell+1\right)} \right]
+ \veps \cal{G} \right\} z_5^0 ,
\label{rdz61dr}
\eeqar
where:
\beqar
{\cal A} &=& e^{-2\nu+2\lambda} \left(\alpha_1 - \alpha_2 \right) r^2 m \sigma_0 \varpi , \\
{\cal B} &=& e^{-2\nu} \left[\left(\frac{\rho}{p} + 1 + \alpha_2 \right) r^3 m \sigma_0 \frac{d\varpi}{dr}
- \left[\left(\left(\frac{\rho}{p} + 1\right)\left(\Gamma+1\right) + \alpha_2 \right) r \frac{d\nu}{dr} + \frac{2}{3} \frac{r}{p}
\frac{d\mu}{dr} + 2 \left(\alpha_1 - \alpha_2 \right) \right] r^2 m \sigma_0 \varpi \right] , \\
{\cal C} &=& e^{-2\nu+2\lambda} \left(\frac{\alpha_3}{\alpha_1} - 1 \right) r^2 m \sigma_0 \varpi , \\
{\cal D} &=& e^{-2\nu} \left[\left(\frac{\rho}{p} + 1 - \alpha_1 \right) \frac{r^3 m \sigma_0}{\ell\left(\ell+1\right)}
\frac{d\varpi}{dr}
- \left[\left(\frac{\rho}{p} + 1 - \alpha_1 \right) r \frac{d\nu}{dr} + \frac{r}{p}\frac{d\mu}{dr}
+ 2 \left(\alpha_1-\alpha_2+\alpha_3 + \frac{\alpha_1\alpha_2}{\alpha_3} \right) \right]
\frac{r^2 m \sigma_0 \varpi}{\ell\left(\ell+1\right)} \right] , \\
{\cal E} &=& e^{-2\nu} \left(\frac{\alpha_1}{\alpha_3} - 1 \right) \frac{r^2 m \sigma_0 \varpi}{\ell\left(\ell+1\right)} , \\
{\cal F} &=& e^{-2\nu} \left(- 4 \alpha_1 + \left(2\alpha_3 - \alpha_2 + \frac{\alpha_1 \alpha_2}{\alpha_3}\right) \ell \left(\ell+1\right) \right)
\frac{r^2 m \sigma_0 \varpi}{\ell\left(\ell+1\right)} , \\
{\cal G} &=& e^{-2\nu} 2 \alpha_1 \left(\ell-1\right) \left(\ell+2\right) \frac{r^2 m \sigma_0 \varpi}{\ell\left(\ell+1\right)} .
\eeqar

From equations (\ref{rdz11dr})-(\ref{rdz61dr})
we get the first-order rotational correction of the eigenfrequency, $\sigma_1$,
and the first-order rotational corrections of the eigenfunctions, $z_i^1 \left(i=1 \dots 6\right)$.
Again, the first four of them refer to the interfacial and shear spheroidal modes while the last two
refer to the torsional toroidal modes.
Bars over $\sigma_0, \sigma_1$ and $\varpi$ indicate dimensionless quantities, scaled by $\sqrt{M/R^3}$;
for example $\bar{\sigma}_0 := \sigma_0 / \sqrt{M/R^3}$.
This set of equations reduces to the sets of equations (58)-(61) and (70)-(71) of \cite{LS1996}
in the Newtonian limit.


\subsection{Fluid core} \label{FluidCore}

To describe the oscillations in the fluid-core region, we use the following functions:
\beqar
y_1^j &=& S^j , \\
y_2^j &=& \left(r\frac{d\nu}{dr}\right)^{-1} \frac{\delta p^j}{\rho+\veps p} \stackrel{\veps \rightarrow 0}{=} \frac{\delta p^j}{gr\rho} .
\eeqar
The zeroth and first-order spheroidal radial functions $H^0$ and $H^1$ are given by the relations:
\beqar
H^0 &=& \frac{y_2^0}{c_1 \bar{\sigma}_0^2} , \\
H^1 &=& \frac{y_2^1}{c_1 \bar{\sigma}_0^2}
+ \left\{ \frac{2m\bar{\varpi}}{\bar{\sigma}_0} + \veps {\cal H} \right\} \frac{y_1^0}{\ell\left(\ell+1\right)}
+ \left\{\frac{2m\bar{\varpi}}{\bar{\sigma}_0} \left(\frac{1}{\ell\left(\ell+1\right)}
- \frac{\bar{\sigma}_1}{m\bar{\varpi}} \right)
+ \veps \frac{1}{\ell\left(\ell+1\right)} {\cal I} \right\} \frac{y_2^0}{c_1 \bar{\sigma}_0^2} ,
\eeqar
cf. relations (44) of \cite{YL2002} for nonrotating relativistic stars and relations (81),(86)-(87) of \cite{LS1996}
for rotating Newtonian stars.

To zeroth order in $\Omega$ $\left(j=0\right)$, the system of equations (\ref{xir_tt})-(\ref{xiphi_tt}) with $\mu=0$
can be recasted into a system of two first-order ordinary differential equations for the functions $y_1^0$ and $y_2^0$:
\beqar
r \frac{dy_1^0}{dr} &=&
-\left(3-\frac{V_1}{\Gamma}+\veps U_2 \right) y_1^0
- \left(\frac{V_1}{\Gamma}-\frac{\ell\left(\ell+1\right)}{c_1\bar{\sigma}_0^2} \right) y_2^0 , \label{rdy10dr} \\
r \frac{dy_2^0}{dr} &=&
 \left(e^{2\veps\lambda}c_1 \bar{\sigma}_0^2 + r A_r \right) y_1^0
- \left(U_1 + r A_r \right) y_2^0 , \label{rdy20dr}
\eeqar
where $A_r$ is the relativistic Schwarzschild discriminant:
\beq
A_r = \frac{1}{\rho+\veps p} \frac{d\rho}{dr} - \frac{1}{\Gamma p} \frac{dp}{dr} .
\eeq

Similarly, to first order in $\Omega$ $\left(j=1\right)$, our system for the functions $y_1^1$ and $y_2^1$ is the following:
\beqar
r \frac{dy_1^1}{dr} &=&
-\left(3-\frac{V_1}{\Gamma}+\veps U_2 \right) y_1^1
- \left(\frac{V_1}{\Gamma}-\frac{\ell\left(\ell+1\right)}{c_1\bar{\sigma}_0^2} \right) y_2^1 \non \\
&+& \left\{ \frac{2m\bar{\varpi}}{\bar{\sigma}_0} + \veps {\cal H} \right\} y_1^0
 +  \left\{\left[\frac{2m\bar{\varpi}}{\bar{\sigma}_0}
- \ell \left(\ell+1\right) \frac{2\bar{\sigma}_1}{\bar{\sigma}_0}\right] + 2 \veps {\cal I} \right\} \frac{y_2^0}{c_1 \bar{\sigma}_0^2} ,
\label{rdy11dr}\\
r \frac{dy_2^1}{dr} &=&
 \left(e^{2\veps \lambda}c_1 \bar{\sigma}_0^2 + r A_r \right) y_1^1
- \left(U_1 + r A_r \right) y_2^1 \non \\
&+& 2 e^{2\veps\lambda} c_1 \bar{\sigma}_0 \bar{\sigma}_1 y_1^0
 - \left\{\frac{2m\bar{\varpi}}{\bar{\sigma}_0} + \veps {\cal H} \right\} y_2^0 ,
\label{rdy21dr}
\eeqar
where:
\beqar
{\cal H} &=& \left(\frac{d\varpi}{dr} - 2 \frac{d\nu}{dr} \varpi \right) \frac{rm}{\sigma_0} , \\
{\cal I} &=& e^{-2\nu} r^2 m \sigma_0 \varpi .
\eeqar
In the Newtonian limit $\veps \rightarrow 0$, $V_1 \rightarrow V$, $\varpi \rightarrow \Omega$
and equations (\ref{rdy10dr})-(\ref{rdy20dr}) and (\ref{rdy11dr})-(\ref{rdy21dr})
reduce to equations (82)-(83) and (84)-(85) of \cite{LS1996}, respectively.
In the relativistic nonrotating limit, $\Omega \rightarrow 0$, we are reduced to equations (42)-(43) of \cite{YL2002}.


\subsection{Boundary and normalization conditions} \label{BNConditions}

At the stellar center $\left(r=0\right)$, the eigenfunctions $y_1^j$ and $y_2^j$ must be regular.
By expanding them in appropriate series, $y_i=\sum_0^n \alpha_{i,n} r^n$
and carrying out some algebraic manipulations, we find:
\beq
c_1 \bar{\sigma}_0^2 y_1^j - \ell y_2^j + F^j = 0 ,
\eeq
where:
\beqar
F^0 &=& 0 , \\
F^1 &=& \frac{2m\varpi}{\sigma_0} \left(\frac{\sigma_1}{m\varpi} - \frac{1}{\ell} \right) c_1 \bar{\sigma}_0^2 y_1^0 .
\eeqar

At the stellar surface $\left(r=R\right)$, the Lagrangian perturbation of the pressure must vanish $\left(\Delta p = 0\right)$.
This eventually leads to a simple relation between $y_1^j$ and $y_2^j$, i.e.:
\beq
y_1^j - y_2^j = 0 .
\eeq

At the fluid-solid interfaces, we require continuity of the tractions. This means the jump conditions:
\beqar
z_1^j &=& y_1^j , \\
z_2^j &=& V_1 \left(y_1^j - y_2^j \right) , \\
z_4^j &=& 0 ,
\eeqar
for the spheroidal modes, and:
\beq
z_6^j = 0 ,
\eeq
for the toroidal modes.

Finally, we normalize our zeroth and first-order eigenfunctions by imposing the conditions:
\beqar
y_1^0 &=& 1 , \\
y_1^1 &=& 0 ,
\eeqar
respectively, at the stellar surface.

\section{Results} \label{Results}

\subsection{Neutron star models} \label{NSmodels}

We work with a set of 34 realistic neutron star models:
we combine EoS A \citep{EOS_A}, WFF3 \citep{EOS_WFF3}, APR \citep{EOS_APR} or L \citep{EOS_L} for the fluid core
with EoS DH \citep{EOS_DH} or NV \citep{EOS_NV} for the solid crust
and, for each combination,
we construct a sequence of models,
beginning from a low-mass model of 1.4M$_\odot$
and reaching, in increments of 0.2M$_\odot$,
the maximum-mass model allowed by that EoS.
That maximum-mass model is $\gtrsim$ 1.6, 1.8, 2.2 and 2.6M$_\odot$
for EoS A, WFF3, APR and L, respectively.
Further details about this set of neutron star models
can be found in \citet{SKS2007a} and in references therein.
Here, we add the following pair of fitting formulas:
\beqar
\ln \left(\Delta r/R\right)
&\simeq& -7.95 \left(\pm 0.04\right) M/R -1.28\left(\pm 0.01 \right) , \; \; \; \mbox{for the DH crustal EoS} , \label{DrMR1} \\
\ln \left(\Delta r/R\right)
&\simeq& -7.84 \left(\pm 0.03\right) M/R -1.04\left(\pm 0.01 \right) , \; \; \; \mbox{for the NV crustal EoS} , \label{DrMR2}
\eeqar
which relate stellar compactness $M/R$ with relative crust thickness $\Delta r/R$. It is obvious that this formulae are valid for typical neutron stars with a fluid core and a crust. In the limit where $M/R\to 0$ the star ``solidifies" or else the crust equation of state is valid for the whole star and the above formulae do not reach the correct limit which is just 1. We should point out that a semi-analytic formula has already been derived  by \cite{SA2007}, this formula is valid for a wide range of stellar compactness and has the correct $M/R\to 0$ limit.
In this formula [eq (B6)] there is a free parameter $\alpha$ which should be fixed by the equation of state. \cite{SA2007}, for the crust EoS DH, find  $\alpha\approx 0.019$ for $\Gamma=4/3$ polytropes and $\alpha\approx 0.023$ for the set of realistic equations of state for the fluid core that they have used.  For our stellar models we find $\alpha \approx 0.020$ for the crust EoS DH and $\alpha=0.026$ for the crust EoS NV. 

\subsection{Torsional modes} \label{Torsional_modes}

\begin{figure}
\centering
\includegraphics[width=100mm,angle=-90]{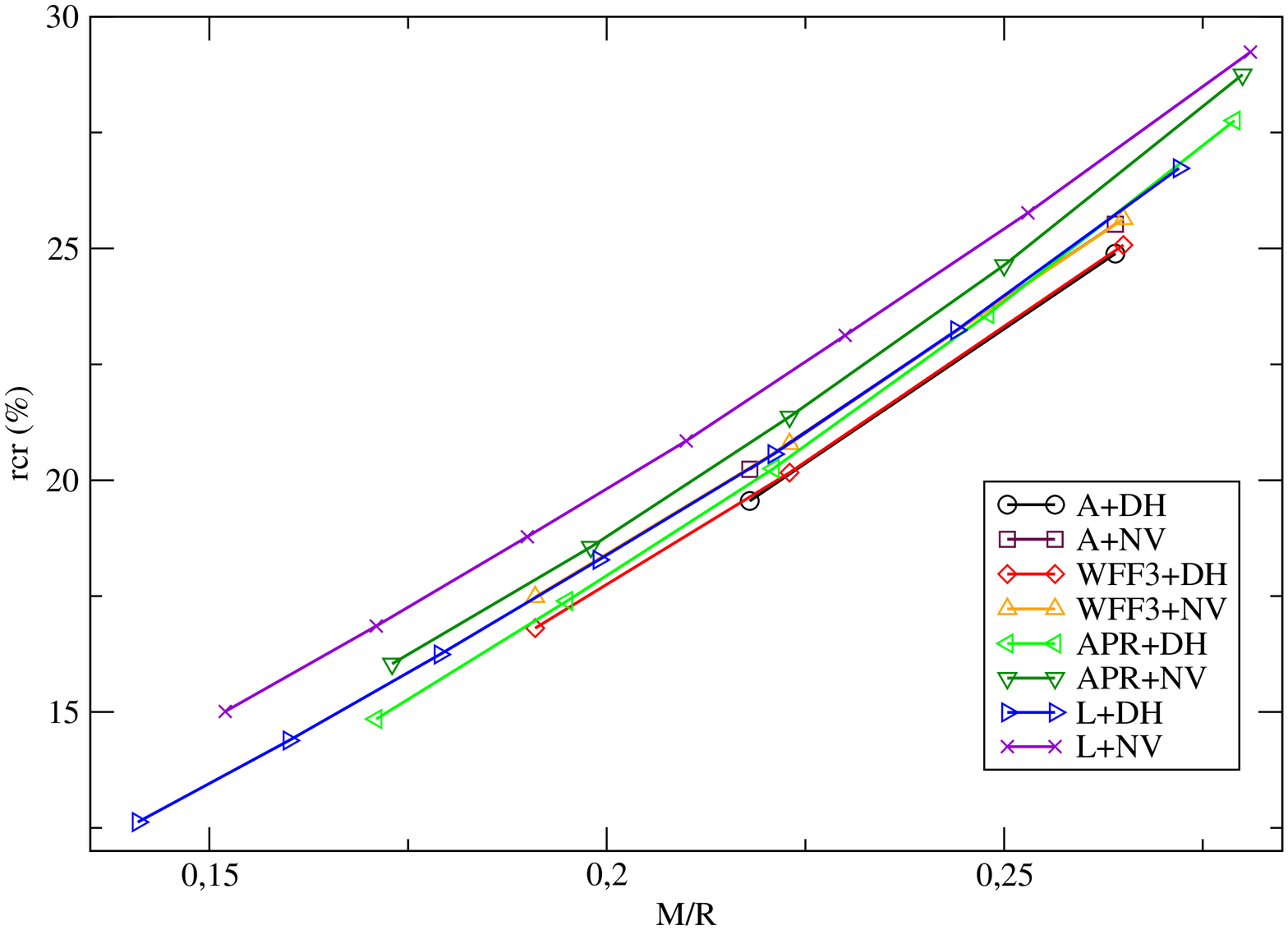}
\caption{
Relativistic correction as measured from a rotating observer $\left(rc_{\mbox{r}}\right)$
versus compactness $\left(M/R\right)$
for the fundamental torsional modes $\left(_\ell t_0\right)$
for various EoS.
}
\label{rcr_MR}
\end{figure}

The Newtonian first-order rotational corrections of the eigenfrequencies of the torsional modes
are shown in the formula of \cite{Strohmayer1991}:
\beq
\sigma=\sigma_0+\sigma_1=\sigma_0 + m \Omega C_{\mbox{1,New}} =\sigma_0 + \frac{m\Omega}{\ell(\ell+1)} ,
\label{Strohmayer_formula}
\eeq
which indicates that $C_{\mbox{1,New}} := \sigma_1/m\Omega$ takes the value $1/\left[\ell\left(\ell+1\right)\right]$
independently of the structure of the star.
For example, $C_{\mbox{1,New}}=0.167$ for $\ell=2$, $C_{\mbox{1,New}}=0.083$ for $\ell=3$, $C_{\mbox{1,New}}=0.05$ for $\ell=4$ and so on.
Moreover, the Newtonian first-order rotational corrections of the eigenfunctions of the torsional modes
can be set equal to zero, $T^1=0$,
cf. \cite{Strohmayer1991} and \cite{LS1996}.
\begin{table}
\centering
\caption{
Eigenfrequencies $\left(\sigma_0\right)$,
Newtonian first-order rotational corrections $\left(C_{\mbox{1,New}}\right)$
and
relativistic first-order rotational corrections $\left(C_{\mbox{1,rel}}\right)$
of the $\ell=2,3$ and 4, $n=0$ {\bf torsional modes} $\left(_2t_0,~_3t_0,~_4t_0\right)$
for various stellar models.
As expected, for each EoS, higher compactnesses $\left(M/R\right)$
result in higher relativistic corrections $\left(rc_{\mbox{r}} := 1-C_{\mbox{1,rel}}/C_{\mbox{1,New}}\right)$.
The Newtonian first-order rotational corrections are equal to $1/\left[\ell\left(\ell+1\right)\right]$
for all stellar models.
}
\begin{tabular}{lccccccccccc}
\hline
&
& & $\ell=2$ &
& & $\ell=3$ &
& & $\ell=4$ &
&
\\
Model & $M/R$ &
$\sigma_0 \left(Hz\right)$ & $C_{\mbox{1,New}}$ & $C_{\mbox{1,rel}}$ &
$\sigma_0 \left(Hz\right)$ & $C_{\mbox{1,New}}$ & $C_{\mbox{1,rel}}$ &
$\sigma_0 \left(Hz\right)$ & $C_{\mbox{1,New}}$ & $C_{\mbox{1,rel}}$ &
$rc_{\mbox{r}} \left(\%\right)$ \\
\hline
A+DH$_{14}$    & 0.218 &   28.4 &  0.167 &  0.134 &   44.9 &  0.083 &  0.067 &   60.3 &   0.05 &  0.040 & 19.55 \\
A+DH$_{16}$    & 0.264 &   27.1 &  0.167 &  0.125 &   42.9 &  0.083 &  0.063 &   57.5 &   0.05 &  0.038 & 24.89 \\
WFF3+DH$_{14}$ & 0.191 &   26.3 &  0.167 &  0.139 &   41.6 &  0.083 &  0.069 &   55.7 &   0.05 &  0.042 & 16.81 \\
WFF3+DH$_{16}$ & 0.223 &   25.2 &  0.167 &  0.133 &   39.8 &  0.083 &  0.067 &   53.4 &   0.05 &  0.040 & 20.16 \\
WFF3+DH$_{18}$ & 0.265 &   24.2 &  0.167 &  0.125 &   38.3 &  0.083 &  0.062 &   51.3 &   0.05 &  0.037 & 25.07 \\
APR+DH$_{14}$  & 0.171 &   24.6 &  0.167 &  0.142 &   38.8 &  0.083 &  0.071 &   52.1 &   0.05 &  0.043 & 14.85 \\
APR+DH$_{16}$  & 0.195 &   23.3 &  0.167 &  0.138 &   36.9 &  0.083 &  0.069 &   49.5 &   0.05 &  0.041 & 17.40 \\
APR+DH$_{18}$  & 0.221 &   22.2 &  0.167 &  0.133 &   35.1 &  0.083 &  0.066 &   47.1 &   0.05 &  0.040 & 20.26 \\
APR+DH$_{20}$  & 0.248 &   21.2 &  0.167 &  0.127 &   33.5 &  0.083 &  0.064 &   45.0 &   0.05 &  0.038 & 23.58 \\
APR+DH$_{22}$  & 0.279 &   20.1 &  0.167 &  0.120 &   31.8 &  0.083 &  0.060 &   42.6 &   0.05 &  0.036 & 27.76 \\
L+DH$_{14}$    & 0.141 &   21.5 &  0.167 &  0.146 &   34.0 &  0.083 &  0.073 &   45.6 &   0.05 &  0.044 & 12.63 \\
L+DH$_{16}$    & 0.160 &   20.5 &  0.167 &  0.143 &   32.5 &  0.083 &  0.071 &   43.6 &   0.05 &  0.043 & 14.38 \\
L+DH$_{18}$    & 0.179 &   19.6 &  0.167 &  0.140 &   31.0 &  0.083 &  0.070 &   41.6 &   0.05 &  0.042 & 16.24 \\
L+DH$_{20}$    & 0.199 &   18.9 &  0.167 &  0.136 &   29.9 &  0.083 &  0.068 &   40.1 &   0.05 &  0.041 & 18.28 \\
L+DH$_{22}$    & 0.221 &   18.1 &  0.167 &  0.132 &   28.7 &  0.083 &  0.066 &   38.5 &   0.05 &  0.040 & 20.56 \\
L+DH$_{24}$    & 0.244 &   17.5 &  0.167 &  0.128 &   27.7 &  0.083 &  0.064 &   37.1 &   0.05 &  0.038 & 23.24 \\
L+DH$_{26}$    & 0.272 &   16.9 &  0.167 &  0.122 &   26.7 &  0.083 &  0.061 &   35.8 &   0.05 &  0.037 & 26.73 \\
&&&&&&&&&&&\\
A+NV$_{14}$    & 0.218 &   28.7 &  0.167 &  0.133 &   45.4 &  0.083 &  0.066 &   60.9 &   0.05 &  0.040 & 20.24 \\
A+NV$_{16}$    & 0.264 &   27.4 &  0.167 &  0.124 &   43.3 &  0.083 &  0.062 &   58.1 &   0.05 &  0.037 & 25.52 \\
WFF3+NV$_{14}$ & 0.191 &   26.7 &  0.167 &  0.138 &   42.2 &  0.083 &  0.069 &   56.6 &   0.05 &  0.041 & 17.48 \\
WFF3+NV$_{16}$ & 0.223 &   25.4 &  0.167 &  0.132 &   40.2 &  0.083 &  0.066 &   53.9 &   0.05 &  0.040 & 20.78 \\
WFF3+NV$_{18}$ & 0.265 &   24.4 &  0.167 &  0.124 &   38.6 &  0.083 &  0.062 &   51.7 &   0.05 &  0.037 & 25.64 \\
APR+NV$_{14}$  & 0.173 &   25.2 &  0.167 &  0.140 &   39.8 &  0.083 &  0.070 &   53.4 &   0.05 &  0.042 & 16.04 \\
APR+NV$_{16}$  & 0.198 &   23.8 &  0.167 &  0.136 &   37.6 &  0.083 &  0.068 &   50.5 &   0.05 &  0.041 & 18.56 \\
APR+NV$_{18}$  & 0.223 &   22.6 &  0.167 &  0.131 &   35.7 &  0.083 &  0.066 &   47.9 &   0.05 &  0.039 & 21.37 \\
APR+NV$_{20}$  & 0.250 &   21.4 &  0.167 &  0.126 &   33.9 &  0.083 &  0.063 &   45.5 &   0.05 &  0.038 & 24.64 \\
APR+NV$_{22}$  & 0.280 &   20.3 &  0.167 &  0.119 &   32.1 &  0.083 &  0.059 &   43.1 &   0.05 &  0.036 & 28.75 \\
L+NV$_{14}$    & 0.152 &   23.2 &  0.167 &  0.142 &   36.6 &  0.083 &  0.071 &   49.2 &   0.05 &  0.043 & 15.01 \\
L+NV$_{16}$    & 0.171 &   21.8 &  0.167 &  0.139 &   34.5 &  0.083 &  0.069 &   46.3 &   0.05 &  0.042 & 16.86 \\
L+NV$_{18}$    & 0.190 &   20.7 &  0.167 &  0.135 &   32.7 &  0.083 &  0.068 &   43.9 &   0.05 &  0.041 & 18.78 \\
L+NV$_{20}$    & 0.210 &   19.7 &  0.167 &  0.132 &   31.1 &  0.083 &  0.066 &   41.8 &   0.05 &  0.040 & 20.85 \\
L+NV$_{22}$    & 0.230 &   18.8 &  0.167 &  0.128 &   29.7 &  0.083 &  0.064 &   39.8 &   0.05 &  0.038 & 23.12 \\
L+NV$_{24}$    & 0.253 &   18.0 &  0.167 &  0.124 &   28.4 &  0.083 &  0.062 &   38.1 &   0.05 &  0.037 & 25.77 \\
L+NV$_{26}$    & 0.281 &   17.2 &  0.167 &  0.118 &   27.2 &  0.083 &  0.059 &   36.5 &   0.05 &  0.035 & 29.24 \\

\hline
\end{tabular}
\label{Torsional0}
\end{table}
The relativistic results can be drawn from Tables \ref{Torsional0} and \ref{Torsional1} and from Figure \ref{rcr_MR}.
The relativistic first-order rotational corrections of the eigenfrequencies of the torsional modes,
as measured from a rotating observer,
are shown in the formula:
\beq
\sigma=\sigma_0+\sigma_1=\sigma_0 + m\Omega C_{\mbox{1,rel}} .
\eeq
Then, the rotating observer compares the Newtonian and the relativistic first-order rotational corrections defining
$rc_{\mbox{r}} := 1-C_{\mbox{1,rel}}/C_{\mbox{1,New}}$.
On the other hand, an inertial observer measures:
\beqar
\sigma&=&\sigma_0 - m \Omega + m \Omega C_{\mbox{1,New}} = \sigma_0 + m \Omega C'_{\mbox{1,New}} , \\
\sigma&=&\sigma_0 - m \Omega + m \Omega C_{\mbox{1,rel}} = \sigma_0 + m \Omega C'_{\mbox{1,rel}} ,
\eeqar
in the Newtonian and in the relativistic case, respectively,
and he defines $rc_{\mbox{i}} := 1-C'_{\mbox{1,rel}}/C'_{\mbox{1,New}}$.
It is easy to check that $rc_{\mbox{r}}$ and $rc_{\mbox{i}}$
obey the simple relation:
\beq
rc_{\mbox{i}} = - \frac{rc_{\mbox{r}}}{1/C_{\mbox{1,New}}-1} .
\label{rcr2rci}
\eeq

In Table \ref{Torsional0} we list the eigenfrequencies $\sigma_0$ and the
relativistic first-order rotational corrections $C_{\mbox{1,rel}}$
for the fundamental $\ell=2,3$ and 4 torsional modes $\left(_2t_0,~_3t_0,~_4t_0\right)$
for our set of 34 stellar models.
The Newtonian first-order rotational corrections $C_{\mbox{1,New}}$ are equal to $1/\left[\ell\left(\ell+1\right)\right]$
for all stellar models according to Strohmayer's formula (\ref{Strohmayer_formula}).
As expected, for each EoS,
higher compactnesses $M/R$
result in higher relativistic corrections $rc_{\mbox{r}}$.
For the chosen set of stellar models,
where $0.14 \lesssim M/R \lesssim 0.28$,
the relativistic corrections $rc_{\mbox{r}}$ vary from 10\% to 30\% approximately.
According to equation (\ref{rcr2rci}) and for $\ell=2$ $\left(C_{\mbox{1,New}}=0.167\right)$,
an inertial observer measures relativistic corrections $rc_{\mbox{i}}=-rc_{\mbox{r}}/5$, that is from -2\% to -6\% approximately.
These results are in good agreement with earlier investigations \citep{VSKB2007}.
\begin{table}
\centering
\caption{
Eigenfrequencies $\left(\sigma_0\right)$,
Newtonian first-order rotational corrections $\left(C_{\mbox{1,New}}\right)$
and
relativistic first-order rotational corrections $\left(C_{\mbox{1,rel}}\right)$
of the $\ell=1$, $n=1,2$ and 3 {\bf torsional modes} $\left(_1t_1,~_1t_2,~_1t_3\right)$
for various stellar models.
As expected, for each EoS, higher compactnesses $\left(M/R\right)$
result in higher relativistic corrections $\left(rc_{\mbox{r}} := 1-C_{\mbox{1,rel}}/C_{\mbox{1,New}}\right)$.
The Newtonian first-order rotational corrections are equal to $1/2$
for all stellar models as $\ell=1$.
}
\begin{tabular}{lcrccrccrccc}
\hline
&
& & $n=1$ &
& & $n=2$ &
& & $n=3$ &
&
\\
Model & $M/R$ &
$\sigma_0 \left(Hz\right)$ & $C_{\mbox{1,New}}$ & $C_{\mbox{1,rel}}$ &
$\sigma_0 \left(Hz\right)$ & $C_{\mbox{1,New}}$ & $C_{\mbox{1,rel}}$ &
$\sigma_0 \left(Hz\right)$ & $C_{\mbox{1,New}}$ & $C_{\mbox{1,rel}}$ &
$rc_{\mbox{r}} \left(\%\right)$ \\
\hline
A+DH$_{14}$    & 0.218 & 1203.9 &    0.5 &  0.406 & 2009.4 &    0.5 &  0.406 & 2775.7 &    0.5 &  0.406 & 18.89 \\
A+DH$_{16}$    & 0.264 & 1527.6 &    0.5 &  0.379 & 2549.4 &    0.5 &  0.379 & 3523.9 &    0.5 &  0.379 & 24.30 \\
WFF3+DH$_{14}$ & 0.191 &  940.7 &    0.5 &  0.419 & 1570.6 &    0.5 &  0.420 & 2165.4 &    0.5 &  0.420 & 16.11 \\
WFF3+DH$_{16}$ & 0.223 & 1099.0 &    0.5 &  0.402 & 1834.0 &    0.5 &  0.403 & 2531.8 &    0.5 &  0.403 & 19.50 \\
WFF3+DH$_{18}$ & 0.265 & 1363.8 &    0.5 &  0.378 & 2275.3 &    0.5 &  0.378 & 3144.0 &    0.5 &  0.378 & 24.48 \\
APR+DH$_{14}$  & 0.171 &  760.0 &    0.5 &  0.429 & 1268.9 &    0.5 &  0.429 & 1748.2 &    0.5 &  0.430 & 14.12 \\
APR+DH$_{16}$  & 0.195 &  858.7 &    0.5 &  0.417 & 1433.5 &    0.5 &  0.417 & 1976.6 &    0.5 &  0.417 & 16.69 \\
APR+DH$_{18}$  & 0.221 &  963.7 &    0.5 &  0.402 & 1608.4 &    0.5 &  0.402 & 2219.7 &    0.5 &  0.403 & 19.59 \\
APR+DH$_{20}$  & 0.248 & 1080.9 &    0.5 &  0.385 & 1803.4 &    0.5 &  0.385 & 2490.8 &    0.5 &  0.386 & 22.95 \\
APR+DH$_{22}$  & 0.279 & 1234.7 &    0.5 &  0.364 & 2060.1 &    0.5 &  0.364 & 2846.4 &    0.5 &  0.365 & 27.17 \\
L+DH$_{14}$    & 0.141 &  529.2 &    0.5 &  0.441 &  884.2 &    0.5 &  0.441 & 1217.4 &    0.5 &  0.442 & 11.83 \\
L+DH$_{16}$    & 0.160 &  585.3 &    0.5 &  0.432 &  977.7 &    0.5 &  0.432 & 1347.9 &    0.5 &  0.433 & 13.61 \\
L+DH$_{18}$    & 0.179 &  647.0 &    0.5 &  0.423 & 1080.6 &    0.5 &  0.423 & 1490.7 &    0.5 &  0.423 & 15.50 \\
L+DH$_{20}$    & 0.199 &  711.2 &    0.5 &  0.412 & 1187.4 &    0.5 &  0.412 & 1639.6 &    0.5 &  0.413 & 17.57 \\
L+DH$_{22}$    & 0.221 &  786.6 &    0.5 &  0.401 & 1313.2 &    0.5 &  0.401 & 1814.0 &    0.5 &  0.401 & 19.88 \\
L+DH$_{24}$    & 0.244 &  872.3 &    0.5 &  0.387 & 1456.0 &    0.5 &  0.387 & 2012.5 &    0.5 &  0.388 & 22.59 \\
L+DH$_{26}$    & 0.272 &  992.6 &    0.5 &  0.369 & 1656.3 &    0.5 &  0.369 & 2290.8 &    0.5 &  0.370 & 26.13 \\
&&&&&&&&&&&\\
A+NV$_{14}$    & 0.218 &  950.5 &    0.5 &  0.402 & 1687.5 &    0.5 &  0.402 & 2390.0 &    0.5 &  0.403 & 19.68 \\
A+NV$_{16}$    & 0.264 & 1190.4 &    0.5 &  0.375 & 2113.3 &    0.5 &  0.375 & 2997.5 &    0.5 &  0.376 & 25.03 \\
WFF3+NV$_{14}$ & 0.191 &  740.2 &    0.5 &  0.415 & 1314.7 &    0.5 &  0.416 & 1860.4 &    0.5 &  0.417 & 16.90 \\
WFF3+NV$_{16}$ & 0.223 &  865.4 &    0.5 &  0.399 & 1536.4 &    0.5 &  0.399 & 2176.3 &    0.5 &  0.400 & 20.24 \\
WFF3+NV$_{18}$ & 0.265 & 1069.5 &    0.5 &  0.374 & 1898.0 &    0.5 &  0.374 & 2691.4 &    0.5 &  0.375 & 25.14 \\
APR+NV$_{14}$  & 0.173 &  615.8 &    0.5 &  0.423 & 1094.2 &    0.5 &  0.423 & 1546.9 &    0.5 &  0.424 & 15.43 \\
APR+NV$_{16}$  & 0.198 &  688.0 &    0.5 &  0.410 & 1222.1 &    0.5 &  0.410 & 1730.2 &    0.5 &  0.411 & 17.98 \\
APR+NV$_{18}$  & 0.223 &  769.0 &    0.5 &  0.396 & 1365.3 &    0.5 &  0.396 & 1934.3 &    0.5 &  0.397 & 20.81 \\
APR+NV$_{20}$  & 0.250 &  857.9 &    0.5 &  0.379 & 1523.0 &    0.5 &  0.380 & 2159.3 &    0.5 &  0.380 & 24.11 \\
APR+NV$_{22}$  & 0.280 &  974.1 &    0.5 &  0.359 & 1728.6 &    0.5 &  0.359 & 2451.9 &    0.5 &  0.359 & 28.26 \\
L+NV$_{14}$    & 0.152 &  483.0 &    0.5 &  0.428 &  858.5 &    0.5 &  0.429 & 1212.3 &    0.5 &  0.430 & 14.35 \\
L+NV$_{16}$    & 0.171 &  524.0 &    0.5 &  0.419 &  931.1 &    0.5 &  0.419 & 1316.6 &    0.5 &  0.420 & 16.21 \\
L+NV$_{18}$    & 0.190 &  567.2 &    0.5 &  0.409 & 1007.6 &    0.5 &  0.409 & 1426.1 &    0.5 &  0.410 & 18.16 \\
L+NV$_{20}$    & 0.210 &  614.5 &    0.5 &  0.399 & 1091.3 &    0.5 &  0.399 & 1545.8 &    0.5 &  0.400 & 20.25 \\
L+NV$_{22}$    & 0.230 &  666.7 &    0.5 &  0.387 & 1183.9 &    0.5 &  0.387 & 1678.1 &    0.5 &  0.388 & 22.55 \\
L+NV$_{24}$    & 0.253 &  728.8 &    0.5 &  0.374 & 1294.0 &    0.5 &  0.374 & 1835.2 &    0.5 &  0.375 & 25.23 \\
L+NV$_{26}$    & 0.281 &  823.8 &    0.5 &  0.356 & 1462.1 &    0.5 &  0.356 & 2074.1 &    0.5 &  0.357 & 28.74 \\

\hline
\end{tabular}
\label{Torsional1}
\end{table}
In Table \ref{Torsional1} we turn our attention to higher $\left(n=1,2,3\right)$ torsional overtones.
Dipole $\left(\ell=1\right)$ modes are now allowed by the angular-momentum conservation law
and we additionally know that these higher overtones are quite insensitive
to the azimuthal harmonic index $\ell$ \citep{HC1980,MVHH1988}.
Table \ref{Torsional1} refers to $\ell=1$
but results for $\ell \gtrsim 2$ are very much alike.
In Appendix \ref{EstProof}, we briefly show that the relativistic first-order rotational corrections
of the eigenfrequencies of the torsional modes
can be estimated also by the integral formula:
\beq
\sigma_1 = m \Omega C_{\mbox{1,rel}}
= m \frac{\int_0^R \varpi \left[1/\Lambda+\veps v_s^2 \left(1-2/\Lambda\right)\right] \left(T^0\right)^2 dr}
{\int_0^R \left(T^0\right)^2 dr} . \label{RCE}
\eeq
The numerical evaluation of formula (\ref{RCE})
and the numerical solution of the previously described eigenvalue problem
yield essentially the same results for $\sigma_1$.

Finally, in Figure \ref{rcr_MR} we show the relativistic corrections $rc_{\mbox{r}}$
versus compactnesses $M/R$ for our set of 34 stellar models,
based on the results of Table \ref{Torsional0}.
Clearly, models with the NV crustal EoS
are shifted towards higher $rc_{\mbox{r}}$
in respect with models with the DH crustal EoS; linear fits reveal that:
\beqar
rc_{\mbox{r}} \left(\%\right) &\simeq& 108.9 \left(\pm 3.0\right) M/R -3.5 \left(\pm 0.7 \right) , \; \; \; \mbox{for the DH crustal EoS} , \\
rc_{\mbox{r}} \left(\%\right) &\simeq& 108.9 \left(\pm 4.9\right) M/R -2.5 \left(\pm 1.1 \right) , \; \; \; \mbox{for the NV crustal EoS} ,
\eeqar
and these can be combined with the fitting formulas (\ref{DrMR1})-(\ref{DrMR2}) of the previous Subsection.

\subsection{Interfacial and shear modes} \label{Interfacial_shear_modes}

The Newtonian first-order rotational corrections
of the eigenfrequencies of the interfacial and shear modes
can not be given by an analytic formula
like Strohmayer's formula (\ref{Strohmayer_formula}).
However, these corrections are given by the integral formula \citep{UnnoEtAl1989,Strohmayer1991,LS1996}:
\beq
C_{\mbox{1,New}} = \frac{\int_0^R \left[2\bar{\xi}^r\bar{\xi}^h + \left(\bar{\xi}^h\right)^2\right] \rho r^2 dr}
{\int_0^R \left[\left(\bar{\xi}^r\right)^2 + \ell\left(\ell+1\right)\left(\bar{\xi}^h\right)^2\right] \rho r^2 dr} ,
\label{C1New}
\eeq
where $\bar{\xi}^r = rS^0,~\bar{\xi}^h = r H^0$
are the zeroth-order radial eigenfunctions of the displacement vector
(compare with equation \ref{xi_spheroidal}).
We clearly see that
the Newtonian first-order rotational corrections
of interfacial and shear spheroidal modes
do depend on the background stellar model
whereas torsional toroidal modes
were proved to be independent.

In Table \ref{Interfacial}
we list values of $C_1$ for the $\ell=1,2$ and 3 interfacial modes
for our set of 34 stellar models.
$C_{\mbox{1,New}}$ is calculated by evaluating the integral formula (\ref{C1New})
while
$C_{\mbox{1,rel}}$ is calculated by solving the eigenvalue problem
(\ref{rdz11dr})-(\ref{rdz41dr}), (\ref{rdy11dr})-(\ref{rdy21dr})
along with the appropriate boundary and jump conditions.
In both cases, we first need to solve
the zeroth-order eigenvalue broblem
in order to determine
the zeroth-order eigenfrequency
and the zeroth-order eigenfunctions.

In Table \ref{Shear} we focus on the shear modes.
These eigenmodes are quite insensitive to the value of $\ell$,
thus we fix $\ell=1$ and
list the Newtonian and the relativistic first-order rotational corrections
of the first three members of this family of modes
$\left(_1s_1,~_1s_2,~_1s_3\right)$.
We note that the eigenfrequencies of the shear modes
are almost equal to those of the higher overtones of the torsional modes.
This is expected as s-modes and t-modes
represent different polarizations of transverse elastic waves.
\begin{table}
\centering
\caption{
Eigenfrequencies $\left(\sigma_0\right)$,
Newtonian first-order rotational corrections $\left(C_{\mbox{1,New}}\right)$
and
relativistic first-order rotational corrections $\left(C_{\mbox{1,rel}}\right)$
of the $\ell=1,2$ and 3 {\bf interfacial modes} $\left(_1i,~_2i,~_3i\right)$
for various stellar models.
}
\begin{tabular}{lccccccccc}
\hline
& \multicolumn{3}{c}{$\ell=1$}
& \multicolumn{3}{c}{$\ell=2$}
& \multicolumn{3}{c}{$\ell=3$}
\\
Model
& $\sigma_0 \left(Hz\right)$ & $C_{\mbox{1,New}}$ & $C_{\mbox{1,rel}}$
& $\sigma_0 \left(Hz\right)$ & $C_{\mbox{1,New}}$ & $C_{\mbox{1,rel}}$
& $\sigma_0 \left(Hz\right)$ & $C_{\mbox{1,New}}$ & $C_{\mbox{1,rel}}$ \\
\hline
A+DH$_{14}$    &   22.5 & 0.485 & 0.393 &   48.0 & 0.152 & 0.125 &   70.5 & 0.069 & 0.059 \\
A+DH$_{16}$    &   19.4 & 0.489 & 0.370 &   43.1 & 0.156 & 0.120 &   63.7 & 0.073 & 0.058 \\
WFF3+DH$_{14}$ &   18.1 & 0.482 & 0.403 &   40.7 & 0.150 & 0.127 &   60.1 & 0.067 & 0.059 \\
WFF3+DH$_{16}$ &   17.7 & 0.486 & 0.390 &   39.4 & 0.153 & 0.125 &   58.2 & 0.070 & 0.059 \\
WFF3+DH$_{18}$ &   17.1 & 0.490 & 0.369 &   38.1 & 0.157 & 0.120 &   56.3 & 0.074 & 0.058 \\
APR+DH$_{14}$  &   19.9 & 0.480 & 0.410 &   42.0 & 0.148 & 0.128 &   61.2 & 0.066 & 0.059 \\
APR+DH$_{16}$  &   17.4 & 0.484 & 0.401 &   37.8 & 0.151 & 0.127 &   55.5 & 0.069 & 0.059 \\
APR+DH$_{18}$  &   16.7 & 0.487 & 0.390 &   36.3 & 0.154 & 0.125 &   53.4 & 0.071 & 0.059 \\
APR+DH$_{20}$  &   18.0 & 0.489 & 0.375 &   37.5 & 0.156 & 0.122 &   54.8 & 0.073 & 0.059 \\
APR+DH$_{22}$  &   15.1 & 0.491 & 0.357 &   32.9 & 0.158 & 0.117 &   48.5 & 0.075 & 0.057 \\
L+DH$_{14}$    &   15.1 & 0.479 & 0.419 &   33.3 & 0.147 & 0.130 &   48.7 & 0.065 & 0.059 \\
L+DH$_{16}$    &   15.9 & 0.482 & 0.413 &   33.9 & 0.150 & 0.130 &   49.5 & 0.067 & 0.060 \\
L+DH$_{18}$    &   13.3 & 0.485 & 0.406 &   29.9 & 0.152 & 0.129 &   44.1 & 0.069 & 0.060 \\
L+DH$_{20}$    &   14.7 & 0.487 & 0.399 &   31.4 & 0.154 & 0.128 &   46.0 & 0.071 & 0.060 \\
L+DH$_{22}$    &   12.5 & 0.488 & 0.389 &   28.1 & 0.156 & 0.126 &   41.4 & 0.073 & 0.060 \\
L+DH$_{24}$    &   13.1 & 0.490 & 0.378 &   28.4 & 0.157 & 0.123 &   41.8 & 0.074 & 0.059 \\
L+DH$_{26}$    &   13.4 & 0.492 & 0.362 &   28.5 & 0.159 & 0.119 &   41.9 & 0.076 & 0.058 \\
&&&&&&&&&\\
A+NV$_{14}$    &   20.7 & 0.480 & 0.385 &   45.4 & 0.147 & 0.121 &   66.7 & 0.065 & 0.056 \\
A+NV$_{16}$    &   19.7 & 0.485 & 0.365 &   43.4 & 0.152 & 0.118 &   63.9 & 0.070 & 0.056 \\
WFF3+NV$_{14}$ &   19.2 & 0.477 & 0.396 &   42.0 & 0.146 & 0.123 &   61.4 & 0.064 & 0.056 \\
WFF3+NV$_{16}$ &   18.3 & 0.482 & 0.384 &   40.1 & 0.150 & 0.122 &   58.9 & 0.067 & 0.057 \\
WFF3+NV$_{18}$ &   17.6 & 0.486 & 0.365 &   38.7 & 0.154 & 0.118 &   56.9 & 0.071 & 0.057 \\
APR+NV$_{14}$  &   18.1 & 0.477 & 0.400 &   39.3 & 0.146 & 0.125 &   57.2 & 0.064 & 0.057 \\
APR+NV$_{16}$  &   17.1 & 0.480 & 0.392 &   37.3 & 0.149 & 0.124 &   54.5 & 0.067 & 0.058 \\
APR+NV$_{18}$  &   16.2 & 0.483 & 0.381 &   35.5 & 0.151 & 0.122 &   52.0 & 0.069 & 0.058 \\
APR+NV$_{20}$  &   15.4 & 0.486 & 0.368 &   33.8 & 0.154 & 0.119 &   49.7 & 0.071 & 0.057 \\
APR+NV$_{22}$  &   14.6 & 0.489 & 0.350 &   32.1 & 0.156 & 0.114 &   47.2 & 0.073 & 0.056 \\
L+NV$_{14}$    &   16.6 & 0.479 & 0.406 &   35.9 & 0.147 & 0.126 &   52.1 & 0.064 & 0.057 \\
L+NV$_{16}$    &   15.7 & 0.482 & 0.400 &   33.9 & 0.149 & 0.126 &   49.4 & 0.067 & 0.058 \\
L+NV$_{18}$    &   14.8 & 0.484 & 0.393 &   32.2 & 0.152 & 0.125 &   47.1 & 0.069 & 0.058 \\
L+NV$_{20}$    &   14.1 & 0.486 & 0.385 &   30.7 & 0.153 & 0.123 &   45.0 & 0.071 & 0.059 \\
L+NV$_{22}$    &   13.4 & 0.488 & 0.375 &   29.4 & 0.155 & 0.122 &   43.1 & 0.072 & 0.058 \\
L+NV$_{24}$    &   12.8 & 0.490 & 0.364 &   28.1 & 0.157 & 0.119 &   41.3 & 0.074 & 0.058 \\
L+NV$_{26}$    &   12.3 & 0.491 & 0.349 &   27.0 & 0.158 & 0.115 &   39.7 & 0.075 & 0.057 \\

\hline
\end{tabular}
\label{Interfacial}
\end{table}
\begin{table}
\centering
\caption{
Eigenfrequencies $\left(\sigma_0\right)$,
Newtonian first-order rotational corrections $\left(C_{\mbox{1,New}}\right)$
and
relativistic first-order rotational corrections $\left(C_{\mbox{1,rel}}\right)$
of the $\ell=1$, $n=1,2$ and 3 {\bf shear modes} $\left(_1s_1,~_1s_2,~_1s_3\right)$
for various stellar models.
}
\begin{tabular}{lrccrccrcc}
\hline
& \multicolumn{3}{c}{$n=1$}
& \multicolumn{3}{c}{$n=2$}
& \multicolumn{3}{c}{$n=3$}
\\
Model
& $\sigma_0 \left(Hz\right)$ & $C_{\mbox{1,New}}$ & $C_{\mbox{1,rel}}$
& $\sigma_0 \left(Hz\right)$ & $C_{\mbox{1,New}}$ & $C_{\mbox{1,rel}}$
& $\sigma_0 \left(Hz\right)$ & $C_{\mbox{1,New}}$ & $C_{\mbox{1,rel}}$ \\
\hline
A+DH$_{14}$    & 1203.7 & 0.492 & 0.399 & 2009.4 & 0.495 & 0.402 & 2775.7 & 0.496 & 0.403 \\
A+DH$_{16}$    & 1527.4 & 0.494 & 0.375 & 2549.4 & 0.497 & 0.376 & 3523.9 & 0.497 & 0.377 \\
WFF3+DH$_{14}$ &  940.6 & 0.490 & 0.411 & 1570.6 & 0.494 & 0.415 & 2165.4 & 0.495 & 0.416 \\
WFF3+DH$_{16}$ & 1098.8 & 0.492 & 0.396 & 1833.9 & 0.496 & 0.399 & 2531.8 & 0.496 & 0.400 \\
WFF3+DH$_{18}$ & 1363.7 & 0.494 & 0.374 & 2275.3 & 0.497 & 0.375 & 3144.0 & 0.497 & 0.376 \\
APR+DH$_{14}$  &  759.8 & 0.488 & 0.419 & 1268.9 & 0.493 & 0.424 & 1748.2 & 0.493 & 0.425 \\
APR+DH$_{16}$  &  858.5 & 0.490 & 0.408 & 1433.4 & 0.494 & 0.412 & 1976.5 & 0.495 & 0.413 \\
APR+DH$_{18}$  &  963.6 & 0.492 & 0.396 & 1608.4 & 0.495 & 0.399 & 2219.7 & 0.496 & 0.400 \\
APR+DH$_{20}$  & 1080.8 & 0.493 & 0.380 & 1803.3 & 0.496 & 0.383 & 2490.8 & 0.497 & 0.383 \\
APR+DH$_{22}$  & 1234.6 & 0.495 & 0.361 & 2060.0 & 0.497 & 0.362 & 2846.4 & 0.497 & 0.363 \\
L+DH$_{14}$    &  529.0 & 0.484 & 0.427 &  884.2 & 0.491 & 0.433 & 1217.4 & 0.491 & 0.434 \\
L+DH$_{16}$    &  585.2 & 0.486 & 0.420 &  977.7 & 0.492 & 0.426 & 1347.9 & 0.493 & 0.427 \\
L+DH$_{18}$    &  646.9 & 0.488 & 0.413 & 1080.5 & 0.494 & 0.417 & 1490.7 & 0.494 & 0.418 \\
L+DH$_{20}$    &  711.1 & 0.490 & 0.404 & 1187.3 & 0.495 & 0.408 & 1639.6 & 0.495 & 0.409 \\
L+DH$_{22}$    &  786.5 & 0.492 & 0.394 & 1313.1 & 0.496 & 0.397 & 1814.0 & 0.496 & 0.398 \\
L+DH$_{24}$    &  872.3 & 0.493 & 0.382 & 1455.9 & 0.496 & 0.384 & 2012.5 & 0.497 & 0.385 \\
L+DH$_{26}$    &  992.5 & 0.495 & 0.366 & 1656.3 & 0.497 & 0.367 & 2290.8 & 0.497 & 0.368 \\
&&&&&&&&&\\
A+NV$_{14}$    &  950.2 & 0.492 & 0.395 & 1687.4 & 0.496 & 0.399 & 2389.9 & 0.496 & 0.400 \\
A+NV$_{16}$    & 1190.2 & 0.494 & 0.371 & 2113.2 & 0.497 & 0.373 & 2997.5 & 0.498 & 0.374 \\
WFF3+NV$_{14}$ &  739.9 & 0.490 & 0.407 & 1314.6 & 0.495 & 0.412 & 1860.3 & 0.495 & 0.413 \\
WFF3+NV$_{16}$ &  865.2 & 0.492 & 0.392 & 1536.3 & 0.496 & 0.396 & 2176.2 & 0.496 & 0.397 \\
WFF3+NV$_{18}$ & 1069.3 & 0.494 & 0.370 & 1898.0 & 0.497 & 0.372 & 2691.3 & 0.498 & 0.373 \\
APR+NV$_{14}$  &  615.5 & 0.488 & 0.413 & 1094.0 & 0.494 & 0.418 & 1546.8 & 0.494 & 0.419 \\
APR+NV$_{16}$  &  687.7 & 0.490 & 0.402 & 1222.0 & 0.495 & 0.406 & 1730.1 & 0.496 & 0.408 \\
APR+NV$_{18}$  &  768.8 & 0.492 & 0.390 & 1365.2 & 0.496 & 0.393 & 1934.2 & 0.496 & 0.394 \\
APR+NV$_{20}$  &  857.7 & 0.493 & 0.375 & 1522.9 & 0.497 & 0.377 & 2159.2 & 0.497 & 0.378 \\
APR+NV$_{22}$  &  973.9 & 0.495 & 0.355 & 1728.5 & 0.502 & 0.360 & 2451.8 & 0.498 & 0.358 \\
L+NV$_{14}$    &  482.6 & 0.486 & 0.416 &  858.3 & 0.493 & 0.422 & 1212.1 & 0.493 & 0.424 \\
L+NV$_{16}$    &  523.7 & 0.488 & 0.408 &  931.0 & 0.494 & 0.414 & 1316.5 & 0.494 & 0.415 \\
L+NV$_{18}$    &  567.0 & 0.489 & 0.400 & 1007.5 & 0.495 & 0.405 & 1426.0 & 0.495 & 0.407 \\
L+NV$_{20}$    &  614.3 & 0.491 & 0.391 & 1091.3 & 0.496 & 0.395 & 1545.7 & 0.496 & 0.397 \\
L+NV$_{22}$    &  666.6 & 0.492 & 0.381 & 1183.8 & 0.496 & 0.385 & 1678.0 & 0.497 & 0.386 \\
L+NV$_{24}$    &  728.7 & 0.494 & 0.369 & 1293.9 & 0.497 & 0.372 & 1835.1 & 0.497 & 0.373 \\
L+NV$_{26}$    &  823.7 & 0.495 & 0.353 & 1462.0 & 0.497 & 0.355 & 2074.1 & 0.498 & 0.356 \\

\hline
\end{tabular}
\label{Shear}
\end{table}

\section{Discussion} \label{Discussion}

In this paper,
we have calculated
the first-order rotational corrections
of the torsional, interfacial and shear modes of realistic neutron stars with crusts
within the framework of General Relativity.
Our results agree, qualitatively, with previous Newtonian results
but differ, quantitatively, a few per cent.
An interesting observation is that,
although the low-frequency torsional modes could not explain all of the 18, 26 and 30 Hz frequencies observed in SGRs,
there is a good chance that one or two of them could be explained by low-frequency interfacial modes
as it can be seen from Table \ref{Interfacial}.
This would be a novel scenario for gravitational-wave detection
since it would imply the excitation of polar-type modes
via their coupling with torsional oscillations.

\begin{figure}
\centering
\includegraphics[width=100mm,angle=-90]{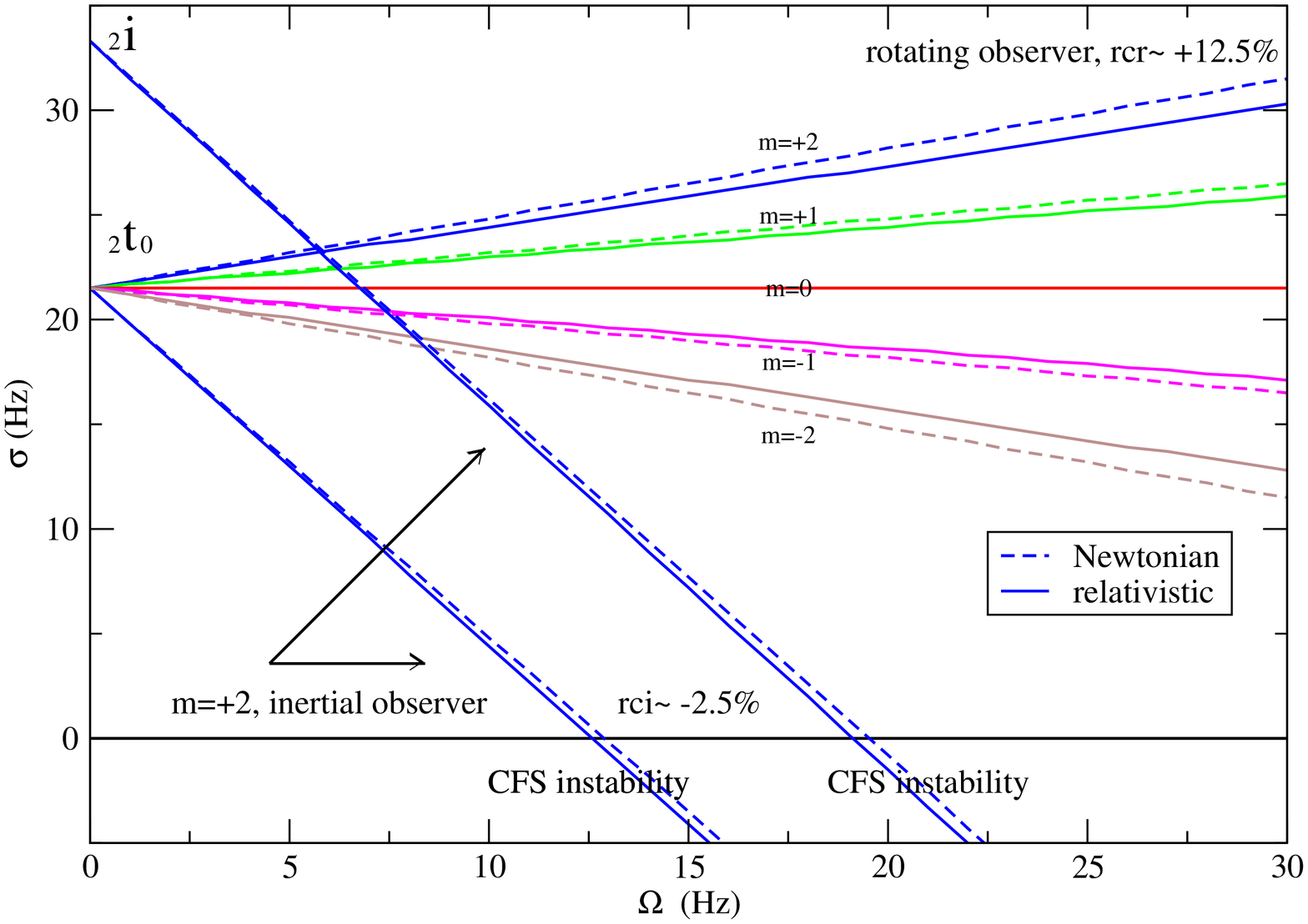}
\caption{
Eigenfrequency $\left(\sigma\right)$ versus
stellar rotational frequency $\left(\Omega\right)$
for the $\ell=2$ interfacial $\left(_2 i\right)$ and fundamental torsional $\left(_2 t_0\right)$ modes
for the stellar model L+DH$_{14}$.
Dashed lines correspond to Newtonian results while solid lines to relativistic ones.
The relativistic corrections are generally small for an inertial observer, $rc_{\mbox{i}} \le rc_{\mbox{r}}$.
However, many low-frequency modes could be driven CFS unstable
and General Relativity shifts the onset of the instability to set in  at lower stellar rotational rates.
}
\label{CFSinst}
\end{figure}

Rotation could drive many oscillation modes unstable
through the Chandrasekhar-Friedman-Schutz (CFS) mechanism.
For an observer rotating with the star,
the instability would set in when 
the eigenfrequency of a mode $\sigma_r$ would be equal to $m\Omega$;
at the same time, the eigenfrequency of the same mode
for a distant inertial observer $\sigma_i$ would be zero.
For the torsional and the interfacial crustal modes
this would happen at moderate stellar rotational frequencies
\citep{YL2001,VSKB2007}.
For example, in Figure \ref{CFSinst},
which refers to the L+DH$_{14}$ EoS,
we show that this would happen at $\Omega \simeq 12.5$Hz
for the torsional $_2t_0$ mode
and at $\Omega \simeq 19$Hz for the interfacial $_2i$ mode.
We have adopted the slow-rotation approximation
but this proves to be sufficient in most cases.
For example, SGRs rotate with periods of $\sim$5-8 secs \citep{WT2006}
which are extremely high compared to their Kepler periods ($\gtrsim$1ms).
On the other hand, for such slow rotators, rotational effects
are quite small; it would be a big surprise
if they could be measured from current x-ray detectors.
Furthermore, CFS instabilities of crustal modes could not be developed in those objects
as they would require stellar rotational frequencies
of several tenths to hundreds of Hz to set in. However, for newly-born neutron stars
things may be quite different. Newly-born neutron stars are expected to rotate much faster,
with frequencies from tenths to hundreds of Hz, and therefore they could become CFS unstable
soon after they would form their crusts. In such a case, the slow-rotation approximation
would be still valid while the effects of rotation in the spectrum
would become pronounced.

\section*{Acknowledgments}
This work is supported by the
Greek GSRT Programs Heracleitus and Pythagoras II and by the German
Science Foundation (DFG) via a SFB/TR7.
A.S. was supported by the National Scholarship Foundation (I.K.Y).


\appendix


\section{Relativistic version of Strohmayer's (1991) equations} \label{StrohmayerGR}

Here we give the explicit form of the three second-order partial differential equations
for the three components of the displacement vector
$\xi^r=\xi^r\left(t,r,\theta,\phi\right)$,
$\xi^\theta=\xi^\theta\left(t,r,\theta,\phi\right)$ and
$\xi^\phi=\xi^\phi\left(t,r,\theta,\phi\right)$.
In order to be able to compare easily with the Newtonian equations (35)-(39) of \citet{Strohmayer1991},
we define the displacement vector as
$\xi^i = \left(\xi^r,\xi^\theta,\xi^\phi \right)
:= \left(\tilde{\xi}^r,\tilde{\xi}^\theta/r,\tilde{\xi}^\phi/r \sin\theta \right)$
and we get:
\beqar
&-& e^{2\veps\left(\lambda-\nu\right)} \frac{\partial^2 \tilde{\xi}^r}{\partial t^2}
+ \veps e^{-2\veps\nu} c_s^2 r \varpi \frac{\partial^2 \tilde{\xi}^\phi}{\partial r \partial t} \sin\theta
\nonumber
\\
&+& e^{-2\veps\nu} \left( \left(2+\veps c_s^2\right) \varpi
+ \veps r \left(\frac{d\varpi}{dr} - \frac{d\nu}{dr} \varpi \right) \left(1+\veps c_s^2\right)
- \veps \Gamma  r \frac{d\nu}{dr} \varpi \right) \frac{\partial \tilde{\xi}^\phi}{\partial t} \sin\theta
\nonumber
\\
&=&
\frac{\partial \chi}{\partial r}
- \frac{A_r \Gamma p}{\rho+\veps p} \alpha
+ \veps c_s^2 \frac{d\nu}{dr} \frac{\partial \tilde{\xi}^r}{\partial r}
+ \veps \left(c_s^2 \left(\frac{d^2\nu}{dr^2} + \left(\frac{d\nu}{dr}\right)^2 \right)
- \left(\Gamma-1\right) \left(\frac{d\nu}{dr}\right)^2 \right) \tilde{\xi}^r
\nonumber
\\
&-&
\frac{\mu}{\rho+\veps p} \left[
\frac{1}{3} \frac{\partial \alpha}{\partial r} + 2 \veps \frac{d\lambda}{dr} \frac{\partial \tilde{\xi}^r}{\partial r}
- \frac{2}{3} \left(\alpha-\veps \left(\frac{d\nu}{dr} + 3 \frac{d\lambda}{dr} \right) \tilde{\xi}^r \right) \frac{1}{\mu} \frac{d\mu}{dr}
+ \frac{2}{\mu} \frac{d\mu}{dr} \frac{\partial \tilde{\xi}^r}{\partial r}
+ e^{2\veps\lambda} \nabla^2 \tilde{\xi}^r \right.
\nonumber
\\
&-&
\left(
\frac{2}{r^2}
+ \veps \frac{1}{3} \frac{d^2\nu}{dr^2}
- \veps \frac{d^2\lambda}{dr^2}
- \veps \frac{4}{3} \frac{d\nu}{dr} \frac{d\lambda}{dr}
+ \veps \frac{4}{3r} \frac{d\nu}{dr}
- \veps \frac{4}{r} \frac{d\lambda}{dr}
\right) \tilde{\xi}^r
- \left(\frac{2}{r^2} + \veps \frac{2}{3r} \frac{d\nu}{dr}\right) \frac{\partial \tilde{\xi}^\theta}{\partial \theta}
\nonumber
\\
&-& \left. \left(\frac{2 \cot\theta}{r^2} + \veps \frac{2\cot\theta}{3r} \frac{d\nu}{dr} \right) \tilde{\xi}^\theta
- \left(\frac{2}{r^2 \sin\theta} + \veps \frac{2}{3r\sin\theta} \frac{d\nu}{dr} \right) \frac{\partial \tilde{\xi}^\phi}{\partial \phi}
\right]
\nonumber
\\
&+&
\veps \frac{\mu}{\rho+\veps p} e^{-2\veps\nu} \left[
\frac{2}{3} r \left(\frac{d\varpi}{dr}
+ \left(\frac{1}{\mu} \frac{d\mu}{dr} - \frac{d\nu}{dr} \right) \varpi \right) \frac{\partial \tilde{\xi}^\phi}{\partial t} \sin\theta
+ \frac{11}{3} \varpi \frac{\partial \tilde{\xi}^\phi}{\partial t} \sin\theta
- \frac{1}{3} r \varpi \frac{\partial^2 \tilde{\xi}^\phi}{\partial r \partial t} \sin\theta
- 2 e^{2\veps\lambda} \varpi \frac{\partial^2 \tilde{\xi}^r}{\partial \phi \partial t}
\right] ,
\label{xir_tt}
\eeqar
\beqar
&-& e^{-2\veps\nu} \frac{\partial^2 \tilde{\xi}^\theta}{\partial t^2}
+ \veps e^{-2\veps\nu} c_s^2 \varpi \frac{\partial^2 \tilde{\xi}^\phi}{\partial \theta \partial t} \sin\theta
+ e^{-2\veps\nu} \left(2 + \veps c_s^2 \right) \varpi \frac{\partial \tilde{\xi}^\phi}{\partial t} \cos\theta
= \frac{1}{r} \frac{\partial \chi}{\partial \theta}
+ \veps c_s^2 \frac{d\nu}{dr} \frac{1}{r} \frac{\partial \tilde{\xi}^r}{\partial \theta}
\nonumber
\\
&-&
\frac{\mu}{\rho+\veps p} \left[
\frac{1}{3r} \frac{\partial \alpha}{\partial \theta}
+ \frac{e^{-2\veps\lambda}}{\mu} \frac{d\mu}{dr} \left(\frac{\partial \tilde{\xi}^\theta}{\partial r}
+ \frac{e^{2\veps\lambda}}{r} \frac{\partial \tilde{\xi}^r}{\partial \theta}
- \frac{\tilde{\xi}^\theta}{r} \right)
+ \nabla^2 \tilde{\xi}^\theta \right.
\nonumber
\\
&+& \left(\frac{2}{r^2} + \veps \frac{2}{3r} \frac{d\nu}{dr} \right) \frac{\partial \tilde{\xi}^r}{\partial \theta}
- \left. \left(e^{-2\veps\lambda}\left(\frac{2}{r^2}+\veps\frac{1}{r}\frac{d\nu}{dr}-\veps\frac{1}{r}\frac{d\lambda}{dr}\right)
-\frac{1}{r^2}+\frac{\cos^2\theta}{r^2 \sin^2\theta} \right) \tilde{\xi}^\theta
- \frac{2 \cos\theta}{r^2 \sin^2\theta} \frac{\partial \tilde{\xi}^\phi}{\partial \phi}
\right]
\nonumber
\\
&+& \veps \frac{\mu}{\rho+\veps p} e^{-2\veps\nu}
\left[\frac{11}{3} \varpi \frac{\partial \tilde{\xi}^\phi}{\partial t} \cos\theta
- \frac{1}{3} \varpi \frac{\partial^2 \tilde{\xi}^\phi}{\partial \theta \partial t} \sin\theta
- 2 \varpi \frac{\partial^2 \tilde{\xi}^\theta}{\partial \phi \partial t} \right] ,
\label{xitheta_tt}
\eeqar
\beqar
&-& e^{-2\veps\nu} \frac{\partial^2 \tilde{\xi}^\phi}{\partial t^2}
- \veps e^{-2\veps\nu} r \varpi \frac{\partial \chi}{\partial t} \sin\theta
+ \veps e^{-2\veps\nu} c_s^2 \varpi \frac{\partial^2 \tilde{\xi}^\phi}{\partial \phi \partial t}
- e^{-2\veps\nu} 2 \varpi \frac{\partial \tilde{\xi}^\theta}{\partial t} \cos\theta
\nonumber
\\
&-& e^{-2\veps\nu} \left(2\varpi + \veps r \left(\frac{d\varpi}{dr} - 2 \frac{d\nu}{dr} \varpi \right)
+ \veps c_s^2 r \frac{d\nu}{dr} \varpi\right)
\frac{\partial \tilde{\xi}^r}{\partial t} \sin\theta
= \frac{1}{r\sin\theta} \frac{\partial \chi}{\partial \phi}
+ \veps c_s^2 \frac{d\nu}{dr} \frac{1}{r\sin\theta} \frac{\partial \tilde{\xi}^r}{\partial \phi}
\nonumber
\\
&-& \frac{\mu}{\rho+\veps p} \left[
\frac{1}{3r\sin\theta} \frac{\partial \alpha}{\partial \phi}
+ \frac{e^{-2\veps\lambda}}{\mu} \frac{d\mu}{dr} \left(\frac{\partial \tilde{\xi}^\phi}{\partial r}
+ \frac{e^{2\veps\lambda}}{r\sin\theta} \frac{\partial \tilde{\xi}^r}{\partial \phi}
- \frac{\tilde{\xi}^\phi}{r} \right)
+ \nabla^2 \tilde{\xi}^\phi \right.
\nonumber
\\
&+& \left(\frac{2}{r^2\sin\theta} + \veps \frac{2}{3r\sin\theta} \frac{d\nu}{dr} \right) \frac{\partial \tilde{\xi}^r}{\partial \phi}
- \left. \left(e^{-2\veps\lambda}\left(\frac{2}{r^2}+\veps\frac{1}{r}\frac{d\nu}{dr}-\veps\frac{1}{r}\frac{d\lambda}{dr}\right)
-\frac{1}{r^2}+\frac{\cos^2\theta}{r^2 \sin^2\theta} \right) \tilde{\xi}^\phi
+ \frac{2 \cos\theta}{r^2 \sin^2\theta} \frac{\partial \tilde{\xi}^\theta}{\partial \phi}
\right]
\nonumber
\\
&+&
\veps \frac{\mu}{\rho+\veps p} e^{-2\veps\nu}
\left[
- r \left(\frac{d\varpi}{dr}
+ \left(\frac{1}{\mu} \frac{d\mu}{dr} - \frac{d\nu}{dr} + \frac{1}{3} \frac{d\lambda}{dr} \right) \varpi \right) \frac{\partial \tilde{\xi}^r}{\partial t} \sin\theta
\right.
- \frac{14}{3} \varpi \frac{\partial \tilde{\xi}^r}{\partial t} \sin\theta
- \frac{13}{3} \varpi \frac{\partial \tilde{\xi}^\theta}{\partial t} \cos\theta
\nonumber
\\
&-& \left. \frac{1}{3} r \varpi \frac{\partial^2 \tilde{\xi}^r}{\partial r \partial t} \sin\theta
- \frac{1}{3} \varpi \frac{\partial^2 \tilde{\xi}^\theta}{\partial \theta \partial t} \sin\theta
- \frac{8}{3} \varpi \frac{\partial^2 \tilde{\xi}^\phi}{\partial \phi \partial t}
\right] ,
\label{xiphi_tt}
\eeqar
where:
\beqar
\chi &:=& - \frac{\Gamma p}{\rho+\veps p} \alpha - \frac{\tilde{\xi}^r}{\rho+\veps p} \frac{dp}{dr}
= - c_s^2 \alpha + \frac{d\nu}{dr} \tilde{\xi}^r , \\
\alpha &:=& \nabla_i \xi^i
= \frac{\partial \tilde{\xi}^r}{\partial r} + \left(\frac{2}{r}
+\veps\frac{d\nu}{dr}+\veps\frac{d\lambda}{dr} \right) \tilde{\xi}^r
+ \frac{1}{r} \frac{\partial \tilde{\xi}^\theta}{\partial \theta} + \frac{\cot\theta}{r} \tilde{\xi}^\theta
+ \frac{1}{r \sin\theta} \frac{\partial \tilde{\xi}^\phi}{\partial \phi} , \\
\nabla^2 &:=& \nabla_i \nabla^i
= e^{-2\veps\lambda} \left[\frac{\partial^2}{\partial r^2}
+ \left(\frac{2}{r}+\veps\frac{d\nu}{dr}-\veps\frac{d\lambda}{dr}\right) \frac{\partial}{\partial r}\right]
+ \frac{1}{r^2} \frac{\partial^2}{\partial \theta^2} + \frac{\cot\theta}{r^2}
\frac{\partial}{\partial \theta}
+ \frac{1}{r^2 \sin^2\theta} \frac{\partial^2}{\partial \phi^2} , \\
\mbox{and where:} \non
\\
c_s^2 &:=& \frac{\Gamma p}{\rho+\veps p} ,
\eeqar
is the speed of sound waves.


\section{Proof of formula (\ref{RCE})} \label{EstProof}

Here we apply a mathematical technique
that allows us to estimate the relativistic first--order rotational corrections
of the eigenfrequencies of the torsional modes through a simple integral formula.
The same technique has been applied in fluid slowly rotating relativistic stars by \citet{YK1997}.

Multiplying equation (\ref{Subequation2}) with the conjugate function $T^{0*}$
and integrating over the whole star we get:
\beqar
\int_0^R
\left\{v_s^2 e^{2\veps\left(\nu-\lambda\right)}
\left[\frac{d^2 T^1}{dr^2}
+\left(\frac{4}{r} + \veps \frac{d\nu}{dr} - \veps \frac{d\lambda}{dr} + \frac{1}{\mu} \frac{d\mu}{dr} \right)
\frac{d T^1}{dr}
- e^{2\veps\lambda} \frac{\Lambda-2}{r^2} T^1 \right]
+ \sigma_0^2 T^1 \right\} T^{0*} dr
\non
\\
+ 2 \sigma_0 \sigma_1 \int_0^R T^0 T^{0*} dr
= 2 m \sigma_0 \int_0^R \varpi \left[\frac{1}{\Lambda}+\veps v_s^2\left(1-\frac{2}{\Lambda}\right)\right] T^0 T^{0*} dr .
\label{EqIntegrated}
\eeqar
But to zeroth order in $\Omega$:
\beq
\int_0^R
\left\{v_s^2 e^{2\veps\left(\nu-\lambda\right)}
\left[\frac{d^2 T^0}{dr^2}
+\left(\frac{4}{r} + \veps \frac{d\nu}{dr} - \veps \frac{d\lambda}{dr} + \frac{1}{\mu} \frac{d\mu}{dr} \right)
\frac{d T^0}{dr}
- e^{2\veps\lambda} \frac{\Lambda-2}{r^2} T^0 \right]
+ \sigma_0^2 T^0 \right\} T^{0*} dr = 0 ,
\eeq
which implies that the first term in equation (\ref{EqIntegrated}) annuls,
leaving the following integral formula for the corrections $\sigma_1$:
\beq
\sigma_1
= m \frac{\int_0^R \varpi \left[1/\Lambda+\veps v_s^2 \left(1-2/\Lambda\right)\right] \left(T^0\right)^2 dr}
{\int_0^R \left(T^0\right)^2 dr} .
\eeq
In the Newtonian limit $\left( \veps \rightarrow 0, \varpi \rightarrow \Omega\right)$,
this formula becomes:
\beq
\sigma_1 = m \Omega \left(1/\Lambda\right) = \frac{m\Omega}{\ell\left(\ell+1\right)} ,
\eeq
as expected.

\label{lastpage}

\end{document}